\documentclass[sigconf]{acmart}
\usepackage{amssymb}

\let\emptyset\varnothing
\usepackage{adjustbox}
\newcommand{\tabincell}[2]{\begin{tabular}{@{}#1@{}}#2\end{tabular}}
\usepackage{listings}
\usepackage{color}
\usepackage{xcolor}
\lstdefinestyle{mystyle}{
    backgroundcolor=\color{backcolour},
    commentstyle=\color{codegreen},
    keywordstyle=\color{magenta},
    numberstyle=\tiny\color{codegray},
    stringstyle=\color{codepurple},
    basicstyle=\scriptsize\ttfamily,
    breaklines=true,
    captionpos=b,
    keepspaces=true,
    numbers=none,
    numbersep=5pt,
    showspaces=false,
    showstringspaces=false,
    showtabs=false,
    tabsize=2
}
\definecolor{codegreen}{rgb}{0,0.6,0}
\definecolor{codegray}{rgb}{0.5,0.5,0.5}
\definecolor{codepurple}{rgb}{0.58,0,0.82}
\definecolor{backcolour}{rgb}{0.95,0.95,0.88}
\usepackage{url}

\setcopyright{none}
\renewcommand{\paragraph}[1]{\vspace{5pt}\noindent\textbf{#1}}

\begin{document}

\title{Good Motive but Bad Design: Why ARM MPU Has Become an Outcast in Embedded Systems}

\author{Wei Zhou}
\affiliation{\textit{National Computer Network Intrusion Protection Center, University of Chinese Academy of Sciences, China}}
\author{Le Guan}
\affiliation{\textit{Department of Computer Science, University of Georgia, USA}}
\author{Peng Liu}
\affiliation{\textit{College of Information Sciences and Technology, The Pennsylvania State University, USA}}
\author{Yuqing Zhang}
\affiliation{\textit{National Computer Network Intrusion Protection Center, University of Chinese Academy of Sciences, China}}

% \affiliation{%

% \affiliation{%


\begin{abstract}
As more and more embedded devices are connected to the Internet, leading to
the emergence of Internet-of-Things (IoT),
previously less tested (and insecure) devices are exposed to miscreants.
To prevent them from being compromised,
% These devices have become hot targets of the hackers.
% Communities are seeking security solutions that work on existing hardware.
% In this context, 
the memory protection unit (MPU),
which is readily available on many devices,
% as a trimmed down version memory management unit (MMU),
has the potential to become a free lunch for the defenders.
% MMU has motivated many novel security systems for our PCs, servers and
% mobile devices.
To our surprise, the MPU is seldom used by real-world
products.
The reasons are multi-fold.
While there are non-technical reasons such as compatibility issues,
more importantly, we found that MPU brings virtually no security enhancement at
the expense of decreased performance and responsiveness.
In this work, we investigate the MPU adoption in major real-time operating systems (RTOSs), in particular, the FreeRTOS,
and try to pinpoint the fundamental reasons to explain why MPU is not favored.
We hope our findings can inspire new remedial solutions to change the situation. 
We also review the latest MPU design and provide
technical suggestions to build more secure embedded systems.
% from the viewpoint of RTOS developers.
\end{abstract}
\maketitle


% \keywords{datasets, neural networks, gaze detection, text tagging}

\section{Introduction}
% \vspace*{-1mm}
Embedded systems have long been operating in a closed environment,
such as industrial plants and vehicle communication systems.
The reliability and robustness of such systems were persistently tested in the past
decades. 
However, these tests were conducted in a benign environment. That is, it is assumed that no adversary could actively penetrate the system.
Unfortunately, this landscape has changed as more and more
embedded devices are exposed to the Internet, where
everyone can launch attack remotely.

Since embedded systems are typically programmed using system programming languages such as C/C++, memory errors pose a great threat to the security
of these systems, especially
considering that many third-party libraries
run at the same privilege level as that of the core program.
% Even worse, some of them (e.g., AmuletOS) allow installation of multiple third-party applications.
On the defense side,
ARM, a leading chip designer for microcontroller unit
(or MCU), proposes the memory protection unit (MPU).
The MPU is a low-cost security extension to ARM MCUs that safeguards certain
sensitive memory regions in case a piece of code is compromised.
% Many other tiny MCUs like the MSP430 FRAM series have also equipped with similar MPU.
Therefore, it is a promising mitigation technique to memory vulnerabilities.
Other MCUs such as MSP430 FRAM followed
this design and implemented similar hardware.
Unfortunately, we found this technique is seldom used in real products,
although the MPU has been around for more than two decades (since the ARMv4t architecture)
% \footnote{\url{https://en.wikipedia.org/wiki/List_of_ARM_microarchitectures}}) 
and has wide adoption on many devices.
Our work is motivated by this observation.
We hope to find out major reasons to explain why the MPU has not been popular.
To do so, we investigate the usage of MPU on embedded operating systems that support it.
This preliminary work reports our results on FreeRTOS,
a leader in the IoT and embedded system market.
We found that the MPU virtually provides no security benefit other
than introducing additional overhead and programming complexity.

% FreeRTOS is a market leader in the IoT and embedded platforms market, being ported to over 40 hardware platforms over the last 14 years. 
% It is a market leading embedded RTOS with over 100,000 downloads per year\cite{FreeRTOSvulns}.
% In November 2017, Amazon Web Services (AWS) took stewardship for the FreeRTOS kernel and its components.

% Most RTOSs lack mitigation protections like W\^X* and ASLR. 
% For instance, withou mpu, any code basd on FreeRTOS is running in supervisor mode.

% multiple vulnerabilities within FreeRTOS's TCP/IP stack and in the AWS secure connectivity modules\cite{}.

% Supporting a model of unprivileged and privileged software execution requires a memory protection scheme that controls the access rights. 
% ARMv7-M supports the Protected Memory System Architecture (PMSAv7). 
% The system address space of a PMSAv7 implementation is protected by a Memory Protection Unit (MPU).

% MMU is used for many functions primarily Virtual Memory (i.e. translation of virtual address to physical address) and memory protection. 
% But MPU is used for memory protection only. 
% In that sense, we can think of MMU as super set of MPU.
\section{The MPU Design}
\label{sec:mpufec}
%++++ how many chips support MPU?++++++++++
% In ARM's portfolio, MCU has always been one of the strongest.
ARM is the leader in MCU design.
% ARM MCUs are mainly comprised of real-time processors (Cortex-R series) and Micro-controller Processors (Cortex-M series),
% both of which are designed to be low-cost and power-efficient,
% with Cortex-R series having an emphasis on hard real-time constraints.
% They implement ARMv7-R and ARMv7-M architectures
% respectively.
% to have a much lower silicon area and
% high-energy efficiency for real-time and industrial applications.
% system line of 
ARM has developed quite a number of different processor products (i.e., Application Processors (Cortex-A series), Real-time Processors (Cortex-R series) and Micro-controller Processors (Cortex-M series)) corresponding to different applications.
Among them, Cortex-R and Cortex-M processors are usually designed to have a much lower silicon area and much high-energy efficiency for real-time and industrial applications.
Therefore, they are very popular in the MCU, deeply embedded systems, and IoT market.

Due to its cost and power-efficient design, ARMv7-M/R have limited security feature support.
% For example, both MMU and W\^{}X are absent.
By default, all the instructions run at the same privilege level
and access the same address space.
As a basic mitigation mechanism,
the MPU has been provided for lightweight access control.
It allows privileged software to define memory regions and assign
memory access permission (read/write/execution)
and memory attributes (ordering and caching) to each of them.
Developers need to configure two registers --
\texttt{Base Address Register (BAR)} and \texttt{Base Attribute/Size Register (BASR)}.
Note that only privileged code can access these registers.
% This design 
% However, to keep high-performance with less silicon area,
% these processors do not have MMU for virtual memory and access control.
% All Cortex-M processors except for Cortex-M0 and M1 based on the ARMv7-M architecture (including its next versions) have an MPU.
% Among them, Cortex-M3 and M4 are most used by current MCUs~\cite{}.
% Unlike an MMU, it
% neither supports virtual memory nor controls memory accesses
% through page tables.
% ARMv7-M processor family supports privilege separation through two separate processor modes, namely privileged and unprivileged modes.
% Based on that, 
% Instead, an MPU provides a fixed number (eight in ARMv7-M)
% of hardware registers,
% and enforces access control rules for each memory region in two permission sets:
% one is used when the processor is in the privileged mode and the other is used when it is in the unprivileged mode.
% It can also define regions of memory whether access is allowed by instruction fetches or used by multiple master at same time.
% The MPU enables setting permissions on regions of physical memory.
% It controls read, write, and execute permissions for both privileged and unprivileged modes.
If a memory access violates the access permissions,
the processor generates a 
% MemManage fault or
HardFault.
The MPU has been supported in main-stream ARM MCUs, including 
Cortex-M0+/M3/M4/M7 and all Cortex-R series.

% The ARMv7-M/R architecture manual describes
% how to use MPU.
The programming model of MPU is described in the ARMv7-M/R architecture manual.
First, depending on the implementation,
an MPU can support 8-16 memory regions
(Cortex-M0+/M3/M4 which are most used by current IoT MCUs only support 8 memory regions\footnote{https://community.arm.com/developer/ip-products/processors/b/processors-ip-blog/posts/arm-cortex-m3-processor-the-core-of-the-iot}).
Each region should be aligned.
Regarding the start address, it must start at an address of multiple of its size.
Regarding the size, it must be 1) at least 32 bytes, and 2) power of two.
Thus, when a region of arbitrary size is required, 
several smaller regions have to be used to reach the target size.
Second, to add flexibly each region greater than 256 bytes can be divided into eight equally sized sub-regions.
This is achieved by configuring the sub-region disable field (SRD) in BASR.
Note that each sub-region can be individually activated, but all still 
have the same permissions.
Third, regions can overlap, and higher numbered regions have precedence. 
There is a default region with priority -1 that maps the entire physical memory.
When enabled, it sets default access permissions for privileged mode.
Unprivileged mode must be explicitly set to grant any permissions.
% Regions can overlap, and higher numbered regions have precedence.

\paragraph{Adoption.}
Taking advantage of the MPU,
embedded systems can restrict access privilege of untrusted or error-prone code,
isolate critical services and therefore mitigate the consequences caused by memory errors.
% For example,
% setting an \emph{execute never (NX)} region can effectively prevent code injection.
Unfortunately, based on our our investigation, the situation is very discouraging.
In Table~\ref{tab:rtos}, we list the adoption of MPU in popular market-leading RTOSs.
ARM Mbed and Nucleus 3.X have integrated the 
MPU in their own system design in the first place.
While others, including FreeRTOS and ThreadX,
add the MPU support optionally.
Other RTOSs such as Contiki, Keil RTX, and TinyOS do not support the MPU at all.
What concerns most is that we found that real devices rarely use MPU-enabled variants.
For example, although FreeRTOS has MPU port,
most manufacturers choose the non-MPU version.
This is evidenced by the fact that Amazon Web Services (AWS), which 
took stewardship of the FreeRTOS kernel in 2017,
only continues to integrate its IoT libraries
to the non-MPU version of FreeRTOS.

% a market leading RTOS and
% and the de-facto standard RTOS for many MUCs,

% We chose several RTOS systems from most widely-used embedded systems which can run on ARM Cortex-M architecture as shown in table~\ref{rtos}
\begin{table}[t]
  \caption{Popular RTOSs MPU adoption}
  \begin{center}
  \begin{adjustbox}{max width=\columnwidth}
  \begin{tabular}{c|c|c}
    \toprule
    \textbf{RTOS}&\textbf{MPU support}&\textbf{Open source}\\
    \midrule
    FreeRTOS&Optional&Open-source\\
    ARM Mbed&Mandatory&Open-source\\
    Nucleus 3.X&Mandatory&Proprietary\\
    Keil RTX&None&Proprietary\\%他网上说叫open source 但其实只是上层接口opensource 下层都是有封装的.a 库TIrtos也是
    Contiki&None&Open-source\\
    ThreadX&None&Proprietary\\
    TinyOS&None&Open-source\\
    TI-RTOS&None&Open-source\\
    uC/OS-II&Optional&Proprietary\\
    VxWorks&Optional&Proprietary\\
    \bottomrule
  \end{tabular}
  \end{adjustbox}
  \end{center} 
\label{tab:rtos}
\end{table}
% Setup a separate privileged data RAM region  
% to protect the kernel data. 

% \vspace*{-4mm}
\section{MPU Usage in FreeRTOS}
% \vspace*{-1mm}
Cooperating with the world's leading chip companies, the FreeRTOS has become a market leading RTOS and the de-facto standard solution for many micro-controllers and small microprocessors in the past 15 years.
Therefore, in this work, we start with studying MPU-enable FreeRTOS.
%In this section, we introduce how an MPU is integrated into the FreeRTOS kernel, a market leading RTOS.
Note that the MPU is only used in 
the MPU-enabled version of FreeRTOS
\footnote{\url{https://www.freertos.org/FreeRTOS-MPU-memory-protection-unit.html}}
rather than the widely adopted normal version.
In the MPU-enabled version,
FreeRTOS is revised in the following ways according to the memory map as shown in Table~\ref{tab:map}.

% We will cover other MPU-enabled RTOSs at the completion of this on-going work.

% ARMv7-M is a memory mapped architecture, meaning that all IO (peripherals and external devices) are directly mapped into the memory space and addressed by dereferencing memory locations. 
% The architecture reserves large amounts of space for each area, but only a small portion of each area is actually used in embedded system.
%According to this memory model,
% To be specific,
% tasks can be created to run in either privileged mode or user mode (i.e, unprivileged mode) and uses the MPU regions to realize the following privilege isolation protections.

1. The code segment was configured to be read-only in case code tampering.
In additional, first 24K of flash is reserved for core FreeRTOS APIs as the second MPU region.
It is set to be read-only in privileged mode, and
inaccessible in unprivileged mode.
%The rest of flash is used to place other code and is readable in unprivileged mode.

2. The kernel maintained data (e.g., current control block) are located in a separated 
512B region in RAM which is readable or writable only in privileged mode by kernel.

3. By default, standard peripherals (e.g., UARTs) could be read or written in unprivileged mode.

4. Tasks can be created to run in either privileged mode or unprivileged mode (user mode).
A separated RAM region is reserved for each user mode task and is inaccessible from any other tasks.
A Privileged mode task can set itself into User mode, but once in User mode it cannot set itself back to Privileged mode.
% Note that this protection is optional, 
% which means the developer can still create regular user mode task without stack isolation but applies other protections.
% The user mode tasks which apply the stack isolation were called restricted use mode tasks. 

5. The remaining three regions can be defined by user mode tasks individually.

6. A memory address like system peripherals (e.g., system timer, Nested Vectored Interrupt Controller (NVIC), MPU registers, etc.) in private peripheral bus (PPB) space which are not mapped by any MPU region falls in the default region,
which can accessed in privileged mode only.

% 6. No data memory is shared between user mode tasks, but user mode tasks can pass messages to each other using the standard queue and semaphore mechanisms. 

% \begin{figure}[t]
% \includegraphics[width=0.5\columnwidth]{memmap}
% \caption{Memory map for ARMv7-M devices}
% \label{fig:memmap}
% \end{figure}

\begin{table}[t]
  \caption{Memory Map of MPU-enabled FreeRTOS on ARMv7-M3/4}
  \begin{center}
  \begin{adjustbox}{max width=\columnwidth}
  \begin{tabular}{c|ccccc}
    \toprule
    Region No.&Base&Size&Usage&\tabincell{c}{Permission\\Mode}&\tabincell{c}{Access\\Attributes}\\
    \midrule
    0&0x0&2GB&Code Segment&\tabincell{c}{Privilege\\User}&\tabincell{c}{rx\\rx}\\
    \hline
    1&0x0&24kB&Kernel APIs&\tabincell{c}{Privilege\\User}&\tabincell{c}{rx\\$\emptyset$}\\
    \hline
    2&0x20000000&512B&Kernel data&\tabincell{c}{Privilege\\User}&\tabincell{c}{rwx\\rx}\\
    \hline
    3&0x40000000&2GB&Standard Peripherals&\tabincell{c}{Privilege\\User}&\tabincell{c}{rw\\rw}\\
    \hline
    4&Stack Bottom&Stack Depth&User Stack&\tabincell{c}{Privilege\\User}&\tabincell{c}{rwx\\rwx}\\
    \hline
    5-7&User-defined&User-defined&User-defined&\tabincell{c}{Privilege\\User}&User-defined\\
    \bottomrule
  \end{tabular}
  \end{adjustbox}
  \end{center}
 \flushleft
\scriptsize{
Memory regions which are unmapped by any MPU region falls in the default region,
which can accessed in privileged mode only.}
\label{tab:map}
\end{table}
\vspace*{-2mm}

\section{Pitfalls of MPU-enabled FreeRTOS}
\label{sec:pitfalls}
We found that the protection implemented in MPU-enable FreeRTOS is incomplete
and can be bypassed easily. 
Moreover, the introduced overhead makes it
unsuitable in many scenarios.

\paragraph{Vulnerable Memory Isolation}
% First, 
The aforementioned memory isolation can be easily bypassed.
Since FreeRTOS APIs are located in a region that can only be accessed in privileged mode,
a task has to raise its privilege temporarily if it needs to invoke a kernel API.
This is achieved by the \texttt{xPortRaisePrivilege} function (\texttt{vPortResetPrivilege}
to drop the privilege).
For example,
as shown in listing~\ref{lst:malloc},
if a user mode task needs memory from the heap,
it has to invoke \texttt{MPU\_pvPortMalloc} which uses \texttt{xPortRaisePrivilege} function
to switch to privilege mode
before invoking the FreeRTOS kernel function \texttt{pvPortMalloc}.
On the completion of \texttt{pvPortMalloc}, it needs to drop the privilege
by invoking \texttt{vPortResetPrivilege}.
Since firmware binaries embed all the static compiled program as well as FreeRTOS AIPs, 
function \texttt{xPortRaisePrivilege} can be easily located via reverse-engineering.
Once attacker is able to carry out control flow hijacking attack (e.g., by exploiting
a memory error) on any user mode task, 
the hijacked task can directly escalate privilege via invoking \texttt{xPortRaisePrivilege}
and never drop the privilege.
With the escalated privilege, the hijacked task can access any resources on the device.

To verify our observation,
we artificially built a firmware with a stack overflow bug,
and then exploited this bug to launch a classical control flow hijacking attack. 
Specifically, the original return address on the stack is overwritten
by the address of the function \texttt{xPortRaisePrivilege}.
As a result, the executing privilege has been escalated.
% user mode has been successfully switched into  mode.
Combining more sophisticated ROP programming,
we were able to take over the control flow with elevated privilege.
% Furthermore, 
% there are no controls that prevent use mode tasks from creating
% privilege tasks and thus 
This has been verified by our experiments and others~\cite{FreeRTOSexp}.
% as mentioned in previous study~\cite{FreeRTOSexp}.
% We have also successfully proved that on MPS2+ FPGA Prototyping System broad (ARM Cortex-M3 and 4).
Ironically, FreeRTOS intends to leverage MPU to protect kernel code from memory errors.
However, improper protection to \texttt{xPortRaisePrivilege} itself
deconstructs the boundary between the task code and kernel code.
In other words, if there is a memory error,
the added security can be bypassed completely.

\begin{lstlisting}[caption={pvPortMalloc function in MPU-enable FreeRTOS},label=lst:malloc]
void *MPU_pvPortMalloc( size_t xSize )
{
    void *pvReturn;
    BaseType_t xRunningPrivileged = xPortRaisePrivilege();
    pvReturn = pvPortMalloc( xSize );
    vPortResetPrivilege( xRunningPrivileged );
    return pvReturn;
}
\end{lstlisting}

% Furthermore,
% a user mode task usually has to use the FreeRTOS APIs (e.g., calling ``pvPortMalloc'' to apply for memory allocation), but the because FreeRTOS APIs are located in the region that can only be accessed while the in privileged mode.
% Thus, the FreeRTOS provides a series of corresponding wrapper APIs to temporarily switch to privilege mode during calling FreeRTOS APIs, which also gives a chance to attacker to raise the privilege by ROP attack.

%还有一点测试过的没写就是 任务切换的时候 判断任务是不是特权任务的只是tcb 栈上的一个bytes, 而且这个地址很容易推算出来 也可以利用简单的rop把那个bytes写出1级可以提权了。

\paragraph{Conflict with Exiting System Design}
Second, due to MPU integration, some existing mechanisms are diminished or 
even become incompatible.
For example,
user mode tasks cannot use dynamic queues because 
there is no shared memory between any two tasks.
To overcome this, a task has to allocate memory statically and shares it
with the peers by configuring an MPU region.
Note that each peer needs an same MPU region for each queue.
In addition, semaphore is also a special kind of 
queue (Its queue length is one).
As a result,
if a task needs multiple queues or semaphore, MPU resources soon become exhausted (there are only
three free MPU regions for user mode tasks to use).
% \begin{lstlisting}[caption={function in MPU-enable FreeRTOS},label=lst:]
% \end{lstlisting}

\paragraph{Incomplete Protection}
The MPU-enable FreeRTOS is 
coarse-grained and inflexible.
First,
standard peripherals are not protected by default.
Although there are three remaining MPU regions can be configured individually by each user mode task, they are not suitable for protecting several separated small peripheral regions due to the alignment problem and limited number of MPU regions as mentioned in Section~\ref{sec:mpufec}.
For instance, 
the memory map for Audio peripheral on MPS2+ FPGA prototyping system broad (AN386) is 0x40024000-0x40024FFF (16 Bytes).
However, the least length of MPU region is 32 bytes.

Second,
Since FreeRTOS  are located in the region that can only be accessed in privileged mode,
a task has to raise its privilege temporarily if it needs to invoke a kernel API.
On the other hand,
developers have the demand to assign separate access rights to
interrupt handlers~\cite{NXPMPUCOM}.
However, NVIC registers are located in the system peripheral region which can only be
accessed in privilege mode unless MPU is
disabled during interrupt handling.
% \footnote{\url{https://community.nxp.com/message/1144517?commentID=1144517\#comment-1144517}}

% standard peripherals are not also directly protected by
% MPU-enable FreeRTOS.

% Furthermore, 
% although the memory regions that are not covered by MPU regions only allows privilege access,
% it enables all access permissions including read, write, and execute, 
% which also has potential hazards.
% Imagine that 
% a attacker exploits a stack overflow in a task stack,
% he cannot execute the malicious code in its stack due to MPU stack isolation.
% However, he can put malicious code in these un-mapped regions beforehand,
% and uses stack overflow attack to redirect the control flow to malicious code in these regions.
% In addition, the developers have the demand to
% assign separate access rights to
% interrupt handlers\footnote{\url{https://community.nxp.com/message/1144517?commentID=1144517\#comment-1144517}}.
% However, the NVIC registers are located in system peripherals region which can only be fully access in privileged mode unless disabling MPU during the ISR.

\paragraph{Increased Overhead}
The ``protection'' provided by the MPU incurs too much overhead.
This is because each invocation to kernel API has to go through a full privilege switch.
% via the \texttt{xPortRaisePrivilege} and \texttt{vPortResetPrivilege} functions.
Since kernel API is frequently invoked in tasks, 
this poses significant impact on real-time performance.
Our experiment shows that
one thousand privilege switch takes 
3.5ms in average on MPS2+ FPGA prototyping system broad (AN386) with 25MHZ CPU clock frequency. 
In a previous research~\cite{kim2018securing}, similar result was obtained.
In addition, 
the MPU regions of each task is different from each other, so
MPU regions have to be reconfigured during task switch, which will also cause time delay.
% because MPU region size and start address restrictions,
% using MPU region will inevitably cause memory waste and fragmentation.
%need our experiment data?
% \vspace*{-5mm}
\section{Why the MPU Has Become an Outcast}
% \vspace*{-1mm}
There are multiple reasons that make MPU less attractive.
We summarizes the most important ones
based on our observations/experiments.

\paragraph{Non-technical Reasons.}
First, IoT devices are low-cost energy-efficient devices.
If more transistors are reserved for complex security features,
not only the price of SoC could be raised accordingly,
but also increased power consumption rules out many applications
in which thermal design power (TDP) concerns.

Second, as IoT business continues to grow, manufacturers are
facing increased time-to-market pressure.
Although security is a concern, manufacturers tend to
reuse existing code base, which is obsolete and less tested
on the Internet.
At the same time, IoT applications are becoming more and more
% sophisticated.
Moreover, developing new software leveraging MPU may
cause compatibility issues. This is clearly shown in
the case of MPU-eabled FreeRTOS and other
developer forums~\cite{STMPUCOM}.
% \footnote{https://community.st.com/s/article/FAQ-Ethernet-not-working-on-STM32H7x3}
In summary, if existing code works,
few companies are willing to harm the profit by investing on security.

\paragraph{Technical Reasons.}
The MPU is a trimmed down version of MMU. 
Can we directly borrow the
design of MMU to replace MPU?
We believe this is infeasible due to two reasons.
Except for the aforementioned cost issue, MPU actually benefits very little
from advanced features  available on MMU.
For example, paging which is the underlying technology of virtual memory, is used
in a batch of security solutions.
Whereas in MCU, paging is an overkill because the RAM in an MCU never
exceeds several megabyte.
The benefit of virtual memory is substantially diminished.
Supporting paging or not becomes a dilemma and ARM obviously chooses not to support it.

Incorporating security checking for each memory access and
frequent privilege switches (as is done in MPU-enabled FreeRTOS) inevitably
introduce performance overhead.
As shown in Section~\ref{sec:pitfalls}, the performance is so significant that
real time constraints cannot be met in some scenarios~\cite{kim2018securing}.
There are two major sources of additional overhead. First, each task switch requires MPU reconfiguration.
Second, each hardware interrupt or invocation of kernel API require privilege escalation.
This rules out the MPU in many real-time applications.

% \vspace*{-3mm}
\section{Suggestions}
% \vspace*{-1mm}
We propose technical suggestions of building
securer embedded systems and review
the latest secure embedded system design.

\subsection{MPU Revision in ARMv8-M}
ARM has already acknowledged the problems with ARMv7-M MPU by revising
it in ARMv8-M.
However, it only mitigates the problem rather than solving it.
The most noticeable improvements include increased region number (as many as 16)
and more flexible region alignment.
As a result, MPU registers can be used more efficient to meet
the requirement of different region sizes.

Using Start and Limit (end) address to define memory regions simplifies memory region definition and leads to a more efficient use of available memory space. 
As mentioned in Section~\ref{sec:mpufec},
when a region of arbitrary size was required, several smaller regions had to be used to reach the target size in ARMv7-M, while ARMv8-M just need one region.
% In addition, all ARMv8-M processors can support MPU regions up to 16.  
% explain which have been fixed (echo A, B, C)

\subsection{Suggestions}

\paragraph{Better Usage of MPU.}
A serious limitation with MPU is that only limited number of regions are supported.
We found that creatively using sub-region disable field (SRD) in BASR can make MPU more efficient.
As mentioned in Section~\ref{sec:mpufec},
each memory region can be divided into eight sub-regions, which can be
enabled/disabled individual.
Suppose a developer needs to allocate a 5KB region and a 3KB region for two tasks.
Without sub-region, 
the developer has to configure two regions of 8KB and 4KB separately.
There is a waste of 3KB and 1KB memory space correspondingly.
With sub-region,
the same 8KB region can be shared between the two tasks.
Specifically,
when running task one, MPU is configured so that the highest three sub-regions of the 8KB region are disabled.
When running task two, MPU is configured so that the lowest
two sub-regions of the 8KB region are disabled.
In this way, the 8KB memory block is reused by the two tasks
without wasting any memory.

In addition,
it is a common practice that same kinds of peripherals (e.g., UART0 and UART1) have adjacent and same size memory region.
Thus, developer can use a large region to cover adjacent and same size peripherals memory region.
If several nearby peripherals need to be protected (i.e., only can be access in privilege mode),
the developer can just disable these correspond sub-regions when running use mode tasks.
% Note that sub-regions cannot be used on regions of size less than 128 bytes.
% Thus, 
Note that
above approach is still not able to protect  several peripheral memory regions which far from each other.

\paragraph{Software Workaround.}
Before a better MPU is proposed,
on the one hand, we should continue to improve coding quality to avoid program bugs;
one the other hand, we can resort to software solutions.
Some previous researches~\cite{kim2018securing,clements2018aces} propose to
use static analysis and 
recompile the firmware to achieve more effective MPU usage.
For example, MINION~\cite{kim2018securing} uses k-means clustering to group
memory sections having similar access permissions together to minimize the 
number of required MPU regions.
It also configures MPU during task switches to avoid privilege escalation requests.

% NDSS/Usenix papers: combine multiple sections into a MPU region
% Securing Real-Time Microcontroller Systems through Customized Memory View Switching
% ACES: Automatic Compartments for Embedded Systems
% uXOM: Efficient eXecute-Only Memory on ARM Cortex-M 
           
\paragraph{Hardware Retrofit.}
% ARM has already acknowledged the problems with ARMv7-M MPU by revising it in ARMv8-M.
% However, it only mitigates the problem rather than solving it.
% The most noticeable improvements include increased region number (as many as 16)
% and more flexible region alignment.
Although, ARMv8-M provide more powerful MPU design,
it can not fundamentally solve the problems we mentioned in Section~\ref{sec:pitfalls}.
% As a result, MPU registers can be used more efficient to meet
% the requirement of different region sizes.
In the long term, we expect a redesigned architecture that fundamentally
addresses the illustrated insecurity and inflexibility
in a lightweight way.
% and at the same time continues to be cost-efficient.
Along this direction, 
hardware-based solutions have
been proposed~\cite{koeberl2014trustlite, nyman2017cfi}.
The most representative work is TrustLite\cite{koeberl2014trustlite}.
It is a radical hardware redesign
that efficiently implements many novel security 
primitives (e.g., execution-aware MPU, secure loader) for embedded devices.
% They design  which allows a flexible allocation and combination of memory and peripheral I/O regions without burdening the CPU.
% also modified CPU exception engine to enable the preemptive scheduling of trusted tasks by an untrusted OS.
In addition, ARMv8-M architecture extends TrustZone technology to Cortex-M series processors for incremental security enhancement.
In particular, existing embedded software does not need heavy re-engineering
but still benefits from a trusted execution domain.
% \vspace*{-3mm}
\section{Conclusion}
    
This paper attempts to answer the question of why the MPU,
as a ready-to-use security feature for protecting ARM-based
MCUs,
has been largely ignored by both device manufacturers and RTOS communities.
We use MPU-enabled FreeRTOS as a concrete example to showcase how the claimed security
benefits brought by MPU can be bypassed or undermined.
Although FreeRTOS cannot represent all the embedded OSs,
we believe our observations apply to other OSs because
the demonstrated pitfalls root in the fundamental design drawbacks of MPU.
We forecast what future MPU will be like and also provide technical suggestions to
safeguard legacy devices.

\bibliographystyle{ACM-Reference-Format}
\bibliography{sample-base}


\begin{thebibliography}{7}

%%% the Web address.

\ifx \showCODEN    \undefined \def \showCODEN     #1{\unskip}     \fi
\ifx \showDOI      \undefined \def \showDOI       #1{#1}\fi
\ifx \showISBNx    \undefined \def \showISBNx     #1{\unskip}     \fi
\ifx \showISBNxiii \undefined \def \showISBNxiii  #1{\unskip}     \fi
\ifx \showISSN     \undefined \def \showISSN      #1{\unskip}     \fi
\ifx \showLCCN     \undefined \def \showLCCN      #1{\unskip}     \fi
\ifx \shownote     \undefined \def \shownote      #1{#1}          \fi
\ifx \showarticletitle \undefined \def \showarticletitle #1{#1}   \fi
\ifx \showURL      \undefined \def \showURL       {\relax}        \fi
% invisible to TeX
\providecommand\bibfield[2]{#2}
\providecommand\bibinfo[2]{#2}
\providecommand\natexlab[1]{#1}
\providecommand\showeprint[2][]{arXiv:#2}

\bibitem[\protect\citeauthoryear{Clements, Almakhdhub, Bagchi, and
  Payer}{Clements et~al\mbox{.}}{2018}]%
        {clements2018aces}
\bibfield{author}{\bibinfo{person}{Abraham~A Clements},
  \bibinfo{person}{Naif~Saleh Almakhdhub}, \bibinfo{person}{Saurabh Bagchi},
  {and} \bibinfo{person}{Mathias Payer}.} \bibinfo{year}{2018}\natexlab{}.
\newblock \showarticletitle{{ACES: Automatic Compartments for Embedded
  Systems}}. In \bibinfo{booktitle}{\emph{27th USENIX Security}}.
\newblock


\bibitem[\protect\citeauthoryear{Community}{Community}{2019}]%
        {NXPMPUCOM}
\bibfield{author}{\bibinfo{person}{NXP Community}.}
  \bibinfo{year}{2019}\natexlab{}.
\newblock \bibinfo{title}{{MPU advanced features?}}
\newblock
  \bibinfo{howpublished}{\url{https://community.nxp.com/message/1144517?commentID=1144517\#comment-1144517}}.
\newblock


\bibitem[\protect\citeauthoryear{Community}{Community}{2018}]%
        {STMPUCOM}
\bibfield{author}{\bibinfo{person}{ST Community}.}
  \bibinfo{year}{2018}\natexlab{}.
\newblock \bibinfo{title}{{FAQ: Ethernet not working on STM32H7x3}}.
\newblock
  \bibinfo{howpublished}{\url{https://community.st.com/s/article/FAQ-Ethernet-not-working-on-STM32H7x3}}.
\newblock


\bibitem[\protect\citeauthoryear{Kim, Kim, Choi, Gu, Lee, Zhang, and Xu}{Kim
  et~al\mbox{.}}{2018}]%
        {kim2018securing}
\bibfield{author}{\bibinfo{person}{Chung~Hwan Kim}, \bibinfo{person}{Taegyu
  Kim}, \bibinfo{person}{Hongjun Choi}, \bibinfo{person}{Zhongshu Gu},
  \bibinfo{person}{Byoungyoung Lee}, \bibinfo{person}{Xiangyu Zhang}, {and}
  \bibinfo{person}{Dongyan Xu}.} \bibinfo{year}{2018}\natexlab{}.
\newblock \showarticletitle{Securing real-time microcontroller systems through
  customized memory view switching}. In \bibinfo{booktitle}{\emph{NDSS}}.
\newblock


\bibitem[\protect\citeauthoryear{Koeberl, Schulz, Sadeghi, and
  Varadharajan}{Koeberl et~al\mbox{.}}{2014}]%
        {koeberl2014trustlite}
\bibfield{author}{\bibinfo{person}{Patrick Koeberl}, \bibinfo{person}{Steffen
  Schulz}, \bibinfo{person}{Ahmad-Reza Sadeghi}, {and} \bibinfo{person}{Vijay
  Varadharajan}.} \bibinfo{year}{2014}\natexlab{}.
\newblock \showarticletitle{TrustLite: A security architecture for tiny
  embedded devices}. In \bibinfo{booktitle}{\emph{EuroSys}}.
\newblock


\bibitem[\protect\citeauthoryear{Nyman, Ekberg, Davi, and Asokan}{Nyman
  et~al\mbox{.}}{2017}]%
        {nyman2017cfi}
\bibfield{author}{\bibinfo{person}{Thomas Nyman}, \bibinfo{person}{Jan-Erik
  Ekberg}, \bibinfo{person}{Lucas Davi}, {and} \bibinfo{person}{N Asokan}.}
  \bibinfo{year}{2017}\natexlab{}.
\newblock \showarticletitle{{CFI CaRE: Hardware-Supported Call and Return
  Enforcement for Commercial Microcontrollers}}. In
  \bibinfo{booktitle}{\emph{RAID}}. Springer.
\newblock


\bibitem[\protect\citeauthoryear{Sandin}{Sandin}{2016}]%
        {FreeRTOSexp}
\bibfield{author}{\bibinfo{person}{Joel Sandin}.}
  \bibinfo{year}{2016}\natexlab{}.
\newblock \bibinfo{title}{{AWS IoT Developer Guide}}.
\newblock
  \bibinfo{howpublished}{\url{https://speakerdeck.com/jsandin/shmoocon-2016-exploiting-memory-corruption-vulnerabilities-on-the-freertos-operating-system}}.
\newblock


\end{thebibliography}

%\appendix

\end{document}